\documentstyle[aps]{revtex}
\begin{document}
\newcommand{\ea}{{\it et al.}}
\newcommand{\PRB}{{Phys. Rev. B }}
\newcommand{\PRL}{{Phys. Rev. Lett. }}
\newcommand{\bi}{\bibitem}
\newcommand{\be}{\begin{equation}}
\newcommand{\ee}{\end{equation}}
\newcommand{\bea}{\begin{eqnarray}}
\newcommand{\eea}{\end{eqnarray}}
\input epsf
\draft
\title{Femtosecond to picosecond electron-energy relaxation 
 and Fr\"ohlich coupling
 in quantum dots}
\author{K. Kr\'al \and Z. Kh\'as}
\address{Institute of Physics, Academy of Sciences of Czech Republic, 
Na
Slovance 2, \\ CZ-18221 Prague 8, Czech Republic}
\date{\today}
\maketitle
\begin{abstract}
Electron relaxation in quantum dots is studied theoretically in polar
semiconductor materials, with an emphasis put on the 
phonon-bottleneck problem and the
 electron-LO-phonon coupling. The theory is based on multiphonon 
 states of
the electron-phonon system and the 
self-consistent Tamm-Dancoff approximation is
used for the electronic self-energy. Electronic relaxation rate 
is shown
numerically to be  on the scale from hundreds fs to tens ps, for 
electron
energy-level separations being in the broad range from about one 
LO-phonon
energy to about three or four 
optical-phonon energies.
Despite of displaying some resonance features, 
the electronic relaxation rate does not appear to be 
crucially dependent on the quantum dot
size. 
\end{abstract}
\pacs{PACS numbers: 72., 72.10.-d, 72.10.Di, 72.15.Lh}
\section{Introduction}
The semiconductor quantum dots (QD) have attracted  attention
 due to the earlier expectations  \cite{Ledentsov,pub1} concerning  
 their
use in high efficiency semiconductor lasers. These expectations of the high
efficiency were 
based mainly 
on the $\delta-$function like electronic density of states of such
zero-dimensional (0D) structures 
and on an expected narrow gain region. In later experiments the luminescence
 of 
 these 0D structures
was found to decrease with decreasing the lateral size of quantum dots
\cite{pub2,pub3,pub4}. Besides attributing this luminescence 
decrease to
technological difficulties in the process of realization of the 
lateral
confinement for electrons and holes, a different explanation
 was suggested
\cite{pub5}, ascribing the luminescence decrease to the so called
'phonon$-$bottleneck effect' (see ref. \cite{pub6}). 
The phonon-bottleneck
hypothesis was supported by an earlier detailed analysis of the 
electron-phonon
interaction in quantum dots based on Born approximation to 
electron-phonon
scattering \cite{pub7,pub8,pub9,pub10}. 

In experiments, the relaxation of  electronic energy was often measured  
 on
the scale of picoseconds 
\cite{pub11,pub12,pub13,pub14,pub15,pub16,pub17,pub18,pub19,pub20}, 
being therefore
fast 
enough to express doubts about the
existence of the bottleneck effect. A significant dependence of the 
electronic relaxation rate on the quantum-dot size was not reported in 
papers 
\cite{pub11,pub12,pub13,pub14,pub15,pub16,pub17,pub18,pub19,pub20} and 
the relaxation
efficiency was reported to be rather 
independent of the relation between the 
electronic
energy-level separations and the optical-phonon energy \cite{pub17}. 
Basing on the
experimental observations, it has been pointed out that the relaxation 
process of electrons in  quantum dots 
should  be considered as a multiphonon process
\cite{pub8,pub12,pub16,pub18,pub19,pub20}, and that the multiphonon
transitions  should be expected to be the main mechanism
providing the electron-energy relaxation. 

The quantum dot system appears to be relatively simple because the
electronic structure of the bound states, 
unperturbed by the lattice motion, may consist of
only several discrete energy levels. It is well known that the
electron-phonon system of the quantum dot can be exactly diagonalized
when the coupling of the electrons to the lattice vibrations is
restricted to the transverse coupling (see below) terms only
\cite{Mahan}. Another simple and exactly soluble model, with no
dissipation of the electronic energy, is the system with two electronic
energy levels coupled to a single mode of the lattice vibrations
\cite{zim}, with the Hamiltonian formally equivalent to Jaynes
Cummings model \cite{Jaynes}. A similar non-dissipating
electron-phonon system was studied in a one-dimensional system
\cite{med}. Although we are aware of several simplified 
systems in which the
electronic subsystem does not relax the energy, it is not yet clear
which of their properties can be generalized to a little more
complicated realistic systems which we likely meet in the real
quantum dots, namely in the systems, in which the carriers, confined in
the dots, interact with a large number of the lattice vibrational modes,
and in which the electron-phonon coupling is more general than that of
the Jaynes Cummings model. Unless a further progress is made in the
area of solving the electronic relaxation problem in the
zero-dimensional structures by exact methods, the approximative
theoretical approaches are in order. 

The role of the multiphonon states in the formulation of the electron
transport problem in quantum dots was emphasized in papers
\cite{pub6,pub21,pub22,pub23,pub24,pub25}. The need for the
self-consistent treatment of the effect of the "collision broadening" of
the electronic energies in quantum dots was emphasized in references
\cite{pub22,pub23}. These two requirements were recently taken into
account in the study of the electronic spectral density
\cite{pub6,pub24}. Very sharp spectral density peaks were obtained. This
sharpness was interpreted as an indicator of a very long electronic
lifetime, without paying a sufficient attention to the lineshape of the
spectral density features. In our previous paper \cite{pub26} it 
was shown that
the spectral density peaks may diverge to infinity as the inverse of 
the square root
of the energy variable. Although such maxima in the electronic spectral
density function are integrable, the relation of the "width" of such
spectral features to the electronic lifetime may not be as simple as it
may be in the case of the Lorentzian peak shape. The relation of the
electronic spectral densities
to the rate of relaxation of the electronic energy, reported
preliminarily in references \cite{pub26,pub46,pub47}, is treated in the
present paper. 

Besides the electron-LO-phonon coupling, other mechanisms were
considered recently as possible candidates for the explanation of the
experimental data on the fast electronic relaxation in quantum dots. So,
the finite lifetime of the optical phonon was shown to provide an
efficient mechanism of relaxing the severe restrictions imposed by the
energy conservation in the Born approximation upon the electron-LO-phonon
scattering \cite{pub48,pub49}. Also, the ultrafast electron energy
relaxation in quantum dots has been recently
 suggested to be explained by the
interaction of the carriers with the defect states in the quantum dot,
taking into account the lattice relaxation mechanism \cite{pub50}.

It has been shown recently, that the electronic scattering in the
low-dimensional structures should be treated with caution. Namely, the
Born approximation to the electronic scattering, giving good results in
the Monte Carlo semiclassical simulations of the electronic transport
properties of the bulk semiconductor samples, appears sometimes to be
rather insufficient in the quasi-two and quasi-one dimensional
structures like quantum wells and quantum wires \cite{Mosko1,Mosko2}.
These observations provide additional arguments in favor of going beyond
the Born approximation in the electron-phonon scattering in quantum
dots.
Understanding the failure of the Born approximation as an implication of
the multiple reflections of the charge carrier from the boundaries of
the low-dimensional structure, it 
can be expected that such  dimensionality effects should be present in the
zero-dimensional structures like quantum dots. In such a case, the
multiple scattering of electrons should be taken into account and the
multiphonon states of the electron-phonon system of the quantum dot
should be of importance. 

The purpose of this paper is to develop further the theory of 
 the electronic relaxation in quantum dots, basing on the multiple
 phonon scattering of the electrons, considering the  interaction 
  of the
 quantum dot electrons with the longitudinal optical phonons of the bulk
 matrix sample into which the quantum dot is built, and to present the
 numerical results of the electronic relaxation rate in dependence on
 the separation of the electronic energies and on the temperature of the
 lattice. 
\section{The model and Hamiltonian}
Generally,  there may be several electrons 
and several
holes 
in a quantum dot structure. 
Assuming the charge neutrality of the system one can speak 
then about
single-exciton, bi-exciton, etc., states, which are strongly 
influenced by the
carrier-carrier coupling, expected  to be significant in quantum dots 
\cite{koch1,koch2}. The inclusion of the carrier-carrier coupling may 
complicate the analysis of the carrier relaxation in quantum dots, 
although it
may be rather significant from the point of view of a quantitative 
comparison
with experimental data. It is the purpose of this paper to 
concentrate the
attention on the electron-phonon mechanism of the carrier 
relaxation. Therefore,
the simplest model system is chosen to be considered, 
namely that consisting of  a
single electron in the quantum dot, while the presence of holes 
in the valence-band states is
completely ignored. The authors believe that the basic features 
of the fast relaxation
mechanism of the excited carriers in quantum dots, 
based on the
electron-LO-phonon Fr\"ohlich interaction, 
are contained already in this 
relatively simple single-electron model.
The inclusion of the carrier-carrier correlation is 
therefore left to a future
work.								      

In order to simplify  
the  numerical part of the work, 
the quantum well is assumed to be of cubic shape, 
with infinitly high electronic
potential energy outside the well. 
The penetration of the electronic stationary wave functions 
into the potential
barriers is therefore neglected. Confining the electronic 
motion to the lowest
lying electronic states is believed to minimize the effect of this 
infinitly-deep
quantum-well approximation.
The Schr\"odinger equation with the
electronic effective mass inside the well gives a set of solutions.
From these electronic states we consider only two states in the present
model: 
the ground state $\psi_0$, with the
energy $E_0$, which will be put equal zero here, and one excited state
$\psi_1$ belonging to the triply degenerate first excited energy $E_1$. 
This set of two nondegenerate electronic unperturbed eigenstates 
provides a
minimum basis set for considering  the
electron-energy relaxation 
due to electronic
transitions between two electronic states coupled via the 
electron-phonon
interaction.
The complete
electron-phonon Hamiltonian $H$ then reads \cite{pub26}
\be
H=H_0+H_1,
\ee
where
\be
H_0=\sum_{n=0}^1 E_nc^+_nc_n + \sum_{\bf q}E_{LO}b^+_{\bf q} b_{\bf q},
\ee
and $H_1$ is Fr\"ohlich coupling \cite{pub26-1}
\be
H_1=\sum^1_{m,n=0}\sum_{\bf q}A_q \Phi(n,m,{\bf q})
(b_{\bf q}-b^+_{-\bf q}
) c^+_nc_m
\ee
of the electron to the system of the dispersionless 
optical phonons of the
whole bulk of the sample with volume $V$, inside of which 
the quantum dot
is built. In the latter equation $q=\mid{\bf q}\mid$, 
$A_q=(-ie/q) [E_{LO}(
\kappa_{\infty}^{-1}-\kappa_0^{-1})]^{1/2}(2\varepsilon_0V)^{-1/2}$, 
where
$\kappa_{\infty}$ and $\kappa_0$ are high-frequency and 
static dielectric
constants, $\varepsilon_0$ is permittivity of free space, 
$e>0$ is
electronic charge. The operator $b_{\bf q}$ annihilates 
the LO phonon with
wavevector ${\bf q}$, while $c_n$ annihilates electron in state 
$\psi_n$.
The electron-phonon coupling depends on the quantum dot size via the
form-factor
\be
\Phi(n,m,{\bf q})=\int d^3{\bf r}\psi^*_n({\bf r})e^{i{\bf 
qr}}\psi_m ({\bf
r}),
\ee
where the integration spreads over the quantum dot volume. It has 
been
demonstrated earlier that in the case of calculating integral 
quantities
like the electronic relaxation rate, it is rather plausible to 
neglect the
impact, which the interfaces in a quasi-two dimensional 
heterostructure of
GaAs-Al$_x$Ga$_{1-x}$As-type might have on the calculated results
\cite{pub27}. It is assumed that the quantum dot structure is based on
the GaAs-Al$_x$Ga$_{1-x}$As-type heterostructure and that the lattice
dynamics of this structure can be  approximated well enough by that of
the bulk GaAs.
In other
cases, like in CdTe quantum dots dispersed in 
poly(vinyl butyral)(PVB)
\cite{pub28}, the influence of the interface modes 
of the optical lattice
vibrations may deserve a more detailed attention \cite{pub29,pub30}.

Let us remark, that the operator $H_1$ of the electron-phonon coupling
contains two kinds of terms. The terms with $n=m$ express such an
interaction of the electron in the state $n$, in which the electron emits
or absorbs a phonon, but the state of the electron remains unchanged.
These term are called transverse interaction terms. The longitudinal
interaction terms, with  $n \neq m$, are those which lead to a change of the
electronic state upon absorbing or emitting a phonon \cite{pubbape}. 
When the longitudinal terms are omitted, then the remaining Hamiltonian
can be diagonalized exactly \cite{Mahan}. Although 
the longitudinal terms may appear to be the
only important ones for the
electronic scattering process of the electronic  transfer from
the excited state to the lower energy state, with emitting or absorbing
simultaneously the phonon,   both the longitudinal and the
transverse terms will be taken into account in this work. The
interaction operator $H_1$ considered in this work does not have the
form of the Jaynes Cummings operator and the electronic system is
coupled to a large number of the phonon modes, so that approximative
theoretical methods will be used in the following analysis.
\section{Relaxation rate}
In  photoluminescence experiments an electron may be excited by a 
light pulse
from the valence-band states into the conduction band states and 
can be finally prepared in an electronic excited state  in the
quantum dot. 
In the case
of a
time-resolved detection of the luminescence, 
corresponding to the process of
annihilation of the electron with a hole in the valence-band 
states, with
the simultaneous emission of a light photon, the rise time of the
luminescence can be determined experimentally 
\cite{pub12,pub13,pub14}. 
The value of the rise time is limited from below by the time which is
needed for  the transport process in which the electron 
is transferred  from the
excited state to the electronic ground state in the dot, 
from which the electron
recombines with a hole in the valence band states. In terms of the
present simple model of the quantum dot one may say, that in this way the
time-derivative $dN_1/dt$ of the population $N_1=<c^+_1c_1>$ of the 
electronic state
$\psi_1$ can be related to the experimental data.

In the present work,
the latter transfer mechanism of the electron, between the electronic
excited and the ground state,  is supposed to be the
process of the scattering of the electron on the system of the optical
phonons.  
In the model of the quantum
dot, considered here, it is assumed 
that at time $t=0$ the system
is prepared in the state with the single electron occupying 
the excited state only, while
the phonon subsystem is found at equilibrium with temperature $T_L$.

The process of the electronic relaxation can be 
theoretically formulated within
the theory of the nonequilibrium statistical 
operator \cite{pub31}. 
The rapidity of the relaxation of the electronic energy can be 
expressed
with the help of the time  derivative of the population, 
$d<c_1^+c_1>/dt$, of the
electronic excited state $(n=1)$.
In the nonequilibrium statistical operator theory, 
assuming  not too short time scale,  the state of the system is assumed
to be 
described
by a set of quantities $<P_k>$, with the corresponding set
$\{P_1,P_2,\dots,P_k,\dots\}$ of operators, suitable for the description
of the system. The averaging in $<P_k>$ 
is
performed with the nonequilibrium statistical operator. In the case
presently considered the electronic subsystem will be characterized 
by
the mean values $N_1=<c^+_1c_1>$ and $N_0=<c^+_0c_0>$ giving the average
population of the electronic excited state and the ground state,
respectively. The state of the phonon system, assumed to be 
permanently
at the thermodynamic equilibrium, will be characterized by the 
mean
value $<b_{\bf q}^+b_{\bf q}>$ given by the Bose-Einstein 
distribution
function at the lattice temperature $T_L$, independent of time.
Confining our choice of $\{P_m\}$ to $<c^+_1c_1>$, $<c^+_0c_0>$ and 
$<b_{\bf q}^+b_{\bf q}>$, the kinetic period of the evolution of the
nonequilibrium system under study is presumed.

The well-known 
expansion of the nonequilibrium statistical operator in powers 
of the
interaction leads to the well-known  expansion  of the "collision
integral" in the generalized kinetic equation giving the time 
evolution
of the mean value $<P_k>$. The lowest-order (in $H_1$) terms of the
collision term of this
equation then read \cite{pub31}:
\be
\frac{\partial<P_k>}{\partial t}=S^{(0)}_k+S_k^{(1)}+S'^{(2)}_k+\dots,
\ee
where
\be
S^{(0)}_k=\frac{1}{i\hbar}<[P_k,H_0]>^t_q
\ee
and
\be
S^{(1)}_k=\frac{1}{i\hbar}<[P_k,H_1]>^t_q. \label{cohe}
\ee
The term $S'^{(2)}_k$, which is at least of the second order in 
$H_1$,
contains generally a factor expressing the memory properties of the
system \cite{pub31}. In the markovian approximation \cite{pub31},
considered here, it
 simplifies 
to the
following expression:
\be
S^{'(2)}_k=-\frac{1}{\hbar^2}\int^0_{-\infty}dt_1 e^{\varepsilon t_1}
<[H_1(t_1),[H_1,P_k]]>^t_q. \label{mark}
\ee 

The averaging $<\dots>^t_q=Tr(\dots \rho_q(t))$ is 
performed with the so-called
quasi-equilibrium statistical operator $\rho_q$ defined as 
\be
\rho_q(t)=Q^{-1}_q\exp \{-\sum_kF_k(t)P_k\},
\ee
\be
Q_q=Tr \exp \{-\sum_kF_k(t)P_k\},
\ee
in which the quantities $F_k(t)$, playing the role of the intensive
quantites describing the system at equilibrium (e.~g. temperature), 
are
determined from the conditions 
\be
<P_k>_q^t=<P_k>.
\ee
 
Consistently with our choice of the operator set $\{P_1,P_2, ...,P_k,
...\}$, 
the term
$S^{(0)}_k$ is  zero.									   
The coupling of the electronic subsystem to the laser light, which is
not considered here,  may 
lead to the
appearence of the interband electronic polarization, which would be  
then a
manifestation of the coherence between the electronic and light 
systems
\cite{schilp}. 
Similarly, the field of the LO phonons can be coherently coupled 
to the
system of excitations of the electronic subsystem. These effects could
be obtained already from the term $S^{(1)}_k$. The right-hand side of
eq. (7) gives then the average value of the product of two electronic
and one phonon operators. Consistently with our choice of the operator
set  $\{P_k\}$ the term  $S^{(1)}_k$ will be ignorred. Extending 
however
the operator set in a suitable way, the Rabi oscillations, studied recently by
Inoshita et al. \cite{pub24}, could be considered simultaneously  with
the  term $S^{(2)}_k$.  The coherent 
coupling between
the electrons and the LO phonons may play a role especially 
in the range of
short time after the light pulse. This effect will not be 
considered in this work.
The lowest-order part 
of the time
derivative of $<P_k>$ is then given by the formula (\ref{mark}).

In the formula (\ref{mark}), giving the lowest-order markovian 
contribution 
to the relaxation rate $\partial <P_k>/\partial t$, 
the operator $H_1(t_1)$ is the interaction operator $H_1$ expressed 
in the 
Heisenberg representation. 
When approximating this Heisenberg representation 
by the interaction
representation, the formula (\ref{mark}) leads directly to the 
well-known Born
approximation to the collision integral. On the other hand, the
Heisenberg representation of the interaction operator $H_1$ in 
(\ref{mark})
makes it possible to sum partially the terms contributing 
to $S'^{(2)}_k$ up
to infinite orders in $H_1$ \cite{pub31}.

The emphasis will be  put now on the obtaining the kinetic equation  
for the
time evolution of the population $<c^+_1c_1>$ of the upper electronic
state. At first, the commutator $[H_1,c^+_1c_1]$ is calculated. As a
result, several terms are obtained, each being a product of two
electronic and one phonon operators. The average in the 
formula (\ref{mark})
then reads:
\bea
\lefteqn{
<[H_1(t_1),[H_1,c^+_1c_1]]>_q^t
=\sum_{r,s,n,{\bf q},{\bf p}}A_qA_p\Phi(r,s,{\bf p})} \\ &\times &
\left\{  
\right. 
\Phi (n,1,{\bf q})    
 < \left[ (b_{\bf p}(t_1)-b^+_{-{\bf p}}(t_1)) 
c^+_r(t_1)c_s(t_1),
 (b_{\bf q}-b^+_{-{\bf
q}}) c^+_nc_1\right] >^t_q \nonumber \\
&-&\Phi(1,n,{\bf q})<\left[ (b_{\bf p}(t_1)-b_{-{\bf p}}^+(t_1) )
c^+_r(t_1)c_s(t_1), (b_{\bf q}-b^+_{-{\bf q}})c^+_1c_n\right]>^t_q
\left. \right\} \nonumber .
\eea
Writing down the commutators in the latter equation explicitly according
to the definition,  the average  quantity
$\Lambda=<[H_1(t_1),[H_1,c^+_1c_1]]>^t_q$ appears to consist of a
number of terms. Each of these terms  is an average of the product of 
six particle 
operators, three
of them being in the Heisenberg representation.  An example of these terms
contributing to $\Lambda$ is:
\be
\sum_{r,s,n,{\bf p},{\bf q}}A_pA_q
\Phi(r,s,{\bf p})\Phi( 1,n,{\bf q}) <b_{-{\bf p}}^+(t_1) 
c^+_r(t_1) c_s(t_1) b_{\bf q}
c^+_1c_n>^t_q.
\ee
The average value in the latter formula, and in the other terms
contributing to $\Lambda$, will be decoupled according to the following
scheme:
\be
<b_{-{\bf p}}^+(t_1)b_{\bf q}c^+_r(t_1)c_s(t_1)c^+_1c_m>^t_q\\
\approx <b_{-{\bf p}}^+(t_1)b_{\bf q}>^t_q <c^+_r(t_1)c_m>^t_q
<c_s(t_1)c_1^+>^t_q, 
\label{scheme}
\ee
in which only the particle 
operators belonging to different interaction operators are
paired. All the anomalous averages of the type like 
 $<b_{\bf p}^+(t_1)b^+_{\bf q}>$ and
$<c^+_r(t_1)c^+_m>$ are considered to be zero. The latter kind of the
decoupling of the averages of the operator products in the relaxation
rate formula was studied earlier in  connection with the electron
energy collision broadening and the electronic relaxation
\cite{zimja,jaofzim} in bulk samples. It was shown, that the above type of the
decoupling leads to the relaxation rate formula with plausible
thermodynamical properties. Although this decoupling scheme appears to
work well in the bulk structures, it is not obvious whether it is
suitable enough in the presently considered two-level zero-dimensional
system. This decoupling is only an approximation to the full correlation
function (13). The question of the
 validity of this decoupling will not be studied in
this work.

In the above obtained single-particle correlation functions, like
$<c_s(t_1)c^+_1>^t_q$, the diagonal approximation 
\be
<c_s(t_1)c^+_r>^t_q \,\,\,\,\approx \,\,\, 
<c_s(t_1)c_r^+>^t_q \delta_{s,r}
\ee
is assumed to hold. This assumption means that 
the quantum number $n$, indexing the unperturbed electronic states
$\psi_n$, remains a good quantum number and that the effects of the
quantum coherence,
included eventually in terms like $<c_1(t_1)c_0^+>^t_q$,
 do not change considerably, in materials with a rather
weak electron-phonon coupling like GaAs, the resulting picture provided
by the diagonal approximation. 
The way of decoupling the average $\Lambda$ is consistent with the
above-made assumption about the choice of the operators $P_k$ describing
the system under study. Nevertheless, the possible 
influence of the nondiagonal
terms like $<c_1(t_1)c_0^+>^t_q$ 
remains to be verified in a future work.

The phonon correlation functions will be taken into consideration in the
diagonal approximation too. This approximation would not be appropriate
when considering the hot phonon effect with the LO phonons produced in
the area of the quantum dot. The hot-phonon effect will be neglected
here.
Assuming that the phonon one-particle correlation functions like
$<b^+_{\bf q}b_{\bf q}(t_1)>$ are 
invariant under the operation of ${\bf
q} \rightarrow -{\bf q}$, we get:
\bea
\lefteqn{<[H_1(t_1),[H_1,c^+_1c_1]]>^t_q} \\
=&&2Re\sum_{\bf q}\mid A_q\mid ^2\mid \Phi(1,0,{\bf q})\mid^2 
\left\{ \right. (<b_{\bf q}(t_1)b^+_{\bf q}>^t_q 
+<b^+_{\bf q}(t_1)b_{\bf q}>^t_q)
   \nonumber \\
&& \times (<c^+_1(t_1)c_1>^t_q<c_0(t_1)c_0^+>^t_q -<c_1(t_1)c^+_1>^t_q 
<c^+_0(t_1)c_0>^t_q)\left. \right\}, \nonumber 
\eea
where $Re$ denotes the real part of a complex number. The knowledge
of the time $t_1$-dependence  of the
single-particle correlation function in the latter equation would make
it possible to perform the integration over $t_1$ in (\ref{mark}) 
and to get in this way
the rate of change of the population $<c^+_1c_1>^t_q$ as a function of
the time $t$.

In calculating the single-particle correlation function we shall 
proceed
in the following way: The single-particle correlation function will be
determined formally for the system with the Hamiltonian (1), in the
case of the thermodynamic equilibrium at a temperature $T$ and the
obtained functional dependence of the electronic and phonon quantities
on the 
temperature will be formally transferred to the case of electrons and
phonons having different respective temperatures $T_e$ and $T_{L}$. 

In the following the electronic correlation functions will be 
expressed in terms of spectral densities  (see eqs. (\ref{corrf1}), 
(\ref{corrf2}) and (\ref{shu}) in
Appendix A).  The electrons and
phonons will be assumed to have different temperatures. Namely, at time
$t=0$ with $N_1=1$, the phonon distribution function 
$\nu_{LO}=<b^+_{\bf q}b_{\bf q}>$ will be 
given by the 
Bose-Einstein distribution function taken at a temperature $T_{L}$.
In the two-level electronic system, unperturbed by
the electron-phonon interaction, with
the energy levels $E_0$ and $E_1$, the chemical potential is
$\mu=(E_0+E_1)/2$. The population $N_1$ of the excited electronic state
at a temperature $T_e$ is
\be
N_1=\frac{1}{1+\exp ( \frac{E_1-E_0}{2k_BT_e})},
\ee
($k_B$ is Boltzmann constant). For example, 
the state of the electronic subsystem
with one electron in the excited state and with the empty ground state
then corresponds to the limit of $T_e \rightarrow 0_-$. In this way the
population of this two-level system can be formally 
expressed with the help of
the electronic temperature $T_e$.
For the purpose of the present calculation  it is suitable to
 describe the state of the two-level electronic
system in terms of the population of the electronic state rather than in
terms of an electronic temperature.

Considering the process of the electron-energy relaxation as 
a markovian
process and taking into account the decoupling (16), 
the theory of the nonequilibrium statistical operator 
leads finally 
to the following formula for the  rate of change $dN_1/dt$ 
of the population of
the electronic state with $n=1$:
\begin{eqnarray}
\frac{dN_1}{dt}  
=-\frac{2\pi}{\hbar} \alpha_{01} \left[N_1(1-N_0)\left( (1+\nu_{LO}) 
\int^{\infty}_{-\infty}dE\,\sigma_1(E)\sigma_0(E-E_{LO})\right. 
\right.  
\label{rate}
\\ 
\left.
+\nu_{LO}\int^{\infty}_{-\infty}dE\sigma_1(E)\sigma_0(E+E_{LO})
\right) 
\nonumber
\\
- N_0(1-N_1)\left((1+\nu_{LO}) 
\int^{\infty}_{-\infty}dE\sigma_0(E)\sigma_1(E-E_{LO})
\right. \nonumber
\\  \left. \left.
+\nu_{LO})\int^{\infty}_{-\infty}dE \sigma_0(E)
\sigma_1(E+E_{LO})\right)\right] 
. \nonumber
\end{eqnarray}
where $\sigma_0$ and $\sigma_1$ are electronic 
spectral densities (see eq. (\ref{shu})), $\nu_{LO}$ 
is
Bose-Einstein distribution of LO-phonons at temperature $T_{L}$ of 
the lattice.
The constant $\alpha_{mn }$ is defined as
\be
\alpha_{mn}=\sum_{\bf q}\mid A_q\mid^2 \mid \Phi (n,m,{\bf q})\mid^2,
~~~~~~\alpha_{mn}=\alpha_{nm}.
\label{alpha}
\ee

The first two terms in the square brackets in eq. (\ref{rate}) are
the only terms contributing at the  the initial moment of time ($t=0$),
 at
which $N_0=0$. The first term provides the production of phonons, while
the second term has the meaning of decreasing the electronic 
population $N_1$ while
absorbing an LO phonon. 
This second term, with the absorption of a phonon, appears in the formula
from  two reasons: First, the relaxation rate formula (18) does
not contain the energy conservation $\delta$-function as it would appear
in the first order of the time dependent perturbation calculation
formula. Second, the term with the absorption of a phonon depends on the
overlap of the two spectral densities, $\sigma_1$ and $\sigma_0$. This
overlap of $\sigma_1$ and $\sigma_0$  is
determined by the properties of the interacting electron-phonon system,
namely, by the multiple-phonon nature of the eigenstates of this system,
which may allow for the non-zero contribution of this phonon-absorption
term to the relaxation rate at $t=0$. In fact, the presently used
approximation to the electronic self-energy (see below) 
assumes the multiple-phonon 
property of the
 eigenstates of the system. The numerical values of the contribution of
 the phonon-absorption term comes out nevertheless 
 rather small in GaAs (see below). This is
 in agreement with the fact, that the electron-LO-phonon coupling in the
 bulk GaAs is not strong.
The last two terms in the square brackets have
the meaning of a transfer of the electron from the state $n=0$ to the
excited state $n=1$. These terms would play a role at the later stages
of the relaxation, namely at $t>0$, which is however not treated here. 

The electronic self-energy is determined in the self-consistent 
Tamm-Dancoff (or the self-consistent Born)
approximation \cite{pub33}. 
The use of the self-consistent Tamm-Dancoff approximation makes it 
possible to
include  the states in which  the electron is coupled  various numbers of 
phonons and
to go beyond the Born approximation when determining  
the self-energy of this highly singular 
system with
$\delta-$function type unperturbed electronic density of states. 
 It is shown in the Appendix B that after a simplification, concerning
 the electronic distribution function $N_m$, the
Dyson equation of the (retarded) electronic self-energy	
for
the electron in the state $n$ can be  
brought to the following form
\cite{pub26}:
\bea
\lefteqn{M_n(E)=\sum^1_{m=0}\alpha_{nm}}
\label{Dyson} 
\\ \times
&\{&\frac{1-N_m+\nu_{LO}}{E-E_m-E_{LO}-M_m(E-E_{LO})+i0_+} 
\nonumber \\
&+&\frac{N_m+\nu_{LO}}{E-E_m+E_{LO}-M_m(E+E_{LO})+i0_+}\},  \nonumber
\eea
where $N_m$ is the electronic population of the $m$-th state. 
In the numerical
evaluations below 
it is assumed that the system is 
at the initial state with $N_1=1$ and $N_0=0$. The
above Dyson equation for $M_n(E)$ assumes that the electronic 
Green's function,
being generally a matrix in the index $n$ in the representation 
of the states
$\psi_n$, can be approximated upon taking the nondiagonal terms 
as zero. This
assumption means that we regard the index $n$ to be 
a "good quantum number".
Also this 
means, that we assume that the mixing of the electronic states $\psi_1$ and
$\psi_0$ via the operator $H_1$ does not contribute seriously 
to the leading
terms of the resulting relaxation rate. This approximation to the
electronic Green's functions is in agreement with what was assumed above
concerning the correlation functions. The extent of the validity 
of such 
approximations in a quantum dot structure 
can be decided by a more detailed study. 

The spectral densities $\sigma_n(E)$ can be obtained  (see eq. (\ref{shu})) 
 from  the real and imaginary part of the 
 retarded self-energy $M_n(E)$,  $M_n(E)=R_n(E)-iY_n(E)$, $Y_n\ge 0$, 
 which is
determined by the equation  eq. (\ref{Dyson}). 
The derivation of this equation,
together with the approximations applied, is presented in Appendix B.
The solution of equation (\ref{Dyson}) can be performed  in two ways: 
(i)
after some simplification this equation allows for an analytical
solution, (ii) the full numerical solution of eq. (\ref{Dyson}) can be
performed in the present model.
\section {An analytical solution}
An approximation can be made, under which the Dyson equation
(\ref{Dyson}) can be solved analytically. This approximation consists
of several simplifications, as follows:

The first approximation step is the neglection of the real part of the
self energy. The relaxation rate (\ref{rate}) is proportional to the
product of the spectral densities, which depend significantly on the
imaginary part of the self-energy. It will be seen later, that the
direct comparison with the result of the numerical solution of the full
equation (\ref{Dyson}) shows, that neglecting the real part of
self-energy does conserve important properties of (\ref{Dyson}). 

\begin{figure}
\epsfxsize=13.5cm
\epsfbox{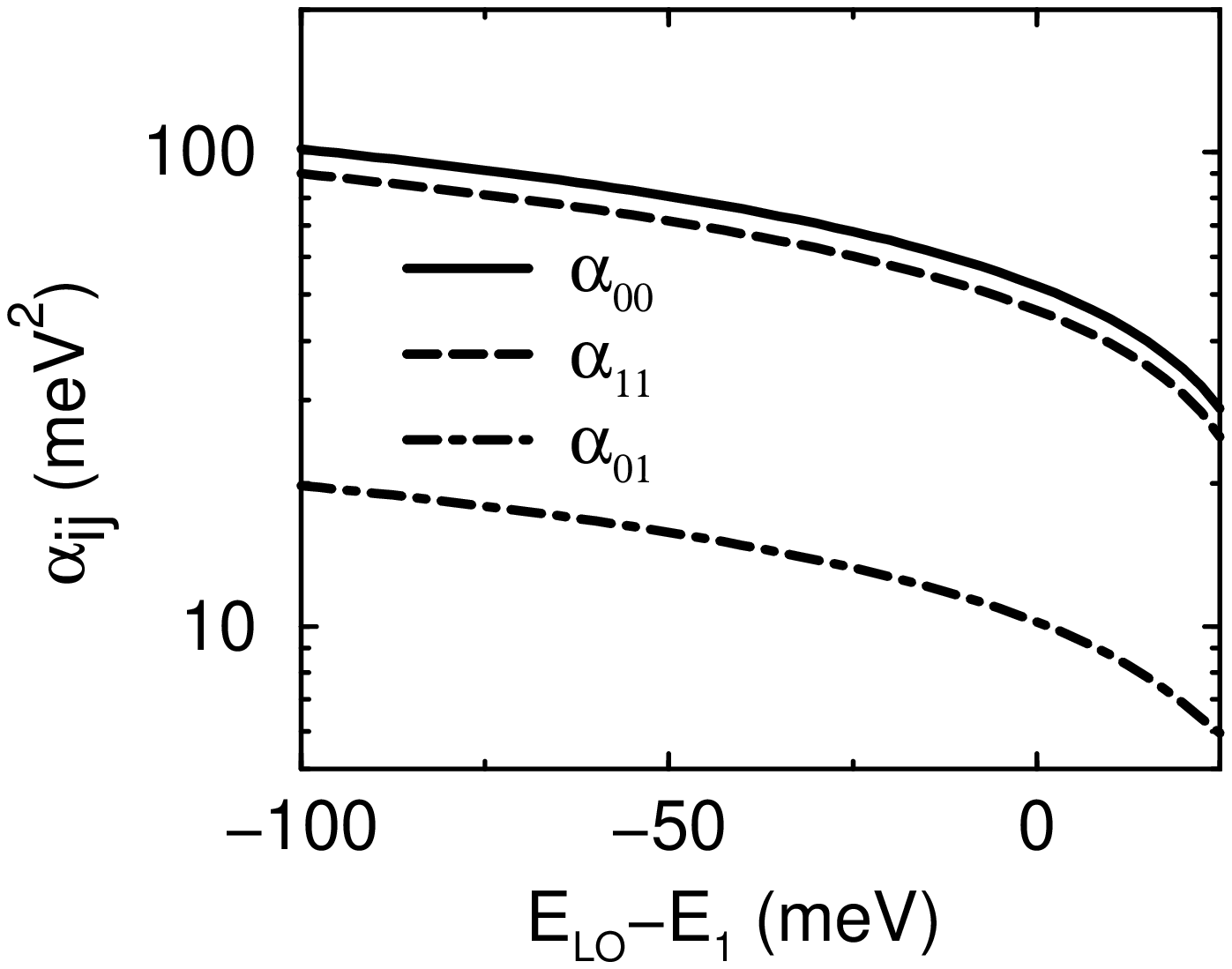}
\caption{The dependence of the coupling constants $\alpha_{ij}$ in GaAs
on the detuning between the optical phonon energy  $E_{LO}$ and the
excited state energy $E_1$.}
\label{graf2ad}
\end{figure}

On the basis of a simple Born approximation approach, it may be expected
that for the relaxation process it  holds approximately that an
electron with energy $E_1$ will make a transition to the state with the
energy $E_0$ and emits an optical phonon with the energy $E_{LO}$ and
that the transverse part of the electron-LO-phonon coupling can
therefore be
neglected. In fact, the constants $\alpha_{00}$ and $\alpha_{11}$
characterizing the transverse coupling are larger than $\alpha_{01}$
  characterizing the longitudinal interaction. The
dependencies of $\alpha_{mn}$ on the detuning between the electronic
excitation energy $E_1$ and the optical phonon energy $E_{LO}$ are
shown in Fig.\ \ref{graf2ad} for GaAs. It
may also be expected  that the imaginary part of the self-energy,
$Y_n(E)$, is significantly nonzero near the energy of $E=E_n$.
Therefore, one of the terms on the rhs of equation (\ref{Dyson}) gives the
leading contribution, while the other term can be neglected. Under these
approximations the Dyson equation (\ref{Dyson}) can be written in the
form of the following set of equations for $Y_n(E)$ (reminding that
$M_n(E)=R_n(E)-iY_n(E)$): 
\be
-Y_1(E)=
\alpha_{01} Im \left\{ \frac{1-N_0+\nu_{LO}}{ E-E_0-E_{LO}+
iY_0(E-E_{LO})+0_+}\right\},
\ee
\be
-Y_0(E)=
\alpha_{01} Im \left\{ \frac{N_1+\nu_{LO}}{ E-E_1+E_{LO}+
iY_1(E+E_{LO})+0_+}\right\}.
\ee
The real part of the self-energy was neglected in this equation.
The non-negative solutions of this equation set are:
\be
Y_0(E)=\frac{\sqrt{\mid E \mid}}{\sqrt{\mid E-E_1+E_{LO}\mid }}
\sqrt{\xi - \mid E \mid \mid E-E_1+E_{LO}\mid},
\ee
\be
Y_1(E)=\frac{\sqrt{\mid E -E_1\mid}}{\sqrt{\mid E-E_{LO}\mid }}
\sqrt{\xi - \mid E -E_1\mid \mid E-E_{LO}\mid},
\ee
where 
 $\xi=\alpha_{01}(1+\nu_{LO})$ and we have  put $E_0$ equal to 
 zero. It is
seen, that the imaginary part of the self-energy $Y_n$ goes to zero
as $\sqrt{E}$ at the energy $E=E_n$. Also, $Y_1$
diverges at $E=E_{LO}$ and $Y_0$ diverges at $E=E_1- E_{LO}$, both like
$1/\sqrt{E}$. The corresponding spectral
densities (see (\ref{shu})) then diverge at these points on the energy
axis like $\sqrt{E}$.

The plot of $Y_0(E)$ and $Y_1(E)$ was presented in the Rapid
Communication \cite{pub26} and will not be repeated here. 
It is
demonstrated there, that 
some peaks 
in $\sigma_n(E)$ of a state with a given $n$ can be understood
as being phonon satellites of the electronic state  with  $m\neq n$. This
property of the self-energy and of the spectral density guarranties, that
the two spectral densities, as they occur in the formula (\ref{rate})
for the relaxation rate, have always a nonzero overlap, no matter how
large the quantum dot is, or, in other words, 
how large the detuning between $E_1-E_0$ and
$E_{LO}$ is.  This property of the self-energy 
can be seen as a formal expression of the idea 
 of the absence of the phonon bottleneck, even in the present
case of electrons interacting with dispersionless phonons. 

\begin{figure}
\epsfxsize=13.5cm
\epsfbox{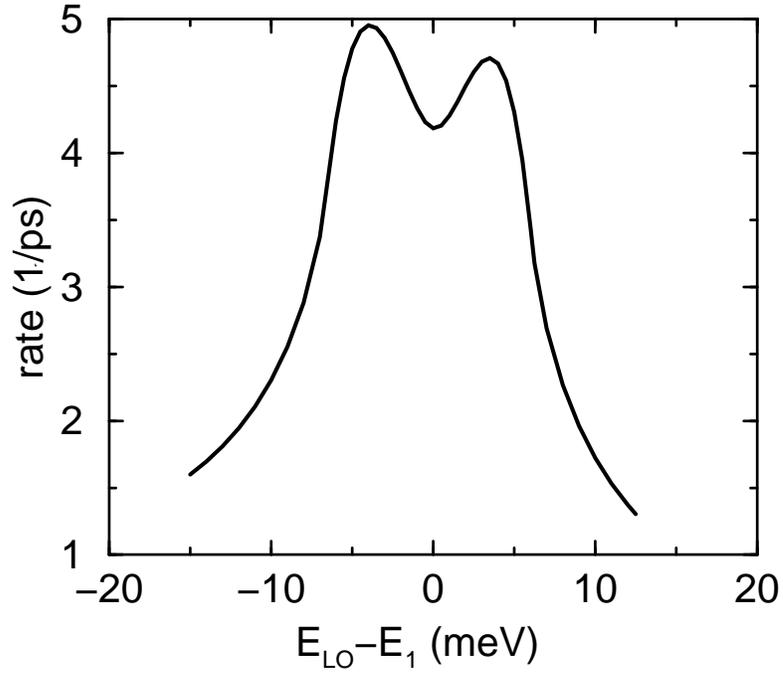}
\caption{The dependence of the electronic relaxation rate $ (-dN_1/dt) $ 
on the detuning   $E_{LO}-E_1$ as computed for the analytical solution
of the simplified Dyson equation (with the transverse coupling terms
neglected) at the  temperatures of the lattice $T_L=77$\,K.}
\label{mtp}
\end{figure}

The relaxation rate $dN_1/dt$ (at $t>0$) computed in the approximation
of the equations (22) and (23) to the electronic self-energy, is shown
in Fig.\ \ref{mtp}. In this
Figure  the relaxation rate  reaches the order of
$1/ps$ in a range of the dot size, in which $E_{LO}-E_1$  varies in the
range of about 30~meV. The relaxation rate in Fig.\ \ref{mtp} demonstrates,
that the present mechanism of the electronic energy relaxation, with the
transverse electron-LO-phonon interaction neglected, can give
the relaxation times in the range of picoseconds or hundreds of
femtoseconds in a rather broad range of the detuning. The double-maximum
shape of the relaxation rate in Fig.\ \ref{mtp} is probably due to the
approximations introduced in the analytical solution and due to the
satellite structure of the electronic spectral densities obtained in
this approximation to the self-energy \cite{pub26}. 

\section{Numerical solution}
In this section the numerical solution of the full equation
(\ref{Dyson}) is presented. That part of the electron-phonon coupling,
which is proportional to the constant $\alpha_{01}$, was shown in
the previous section to give the effect of the absence of the phonon
bottleneck and the relaxation times in the range of picoseconds. 
In the present section the coupling constants $\alpha_{00}$
and $\alpha_{11}$, characterizing the transverse coupling of electrons
to the phonons, are taken into account and 
both the real and imaginary parts
of the self-energy are included into the computation.   

\begin{figure}
\epsfxsize=13.5cm
\epsfbox{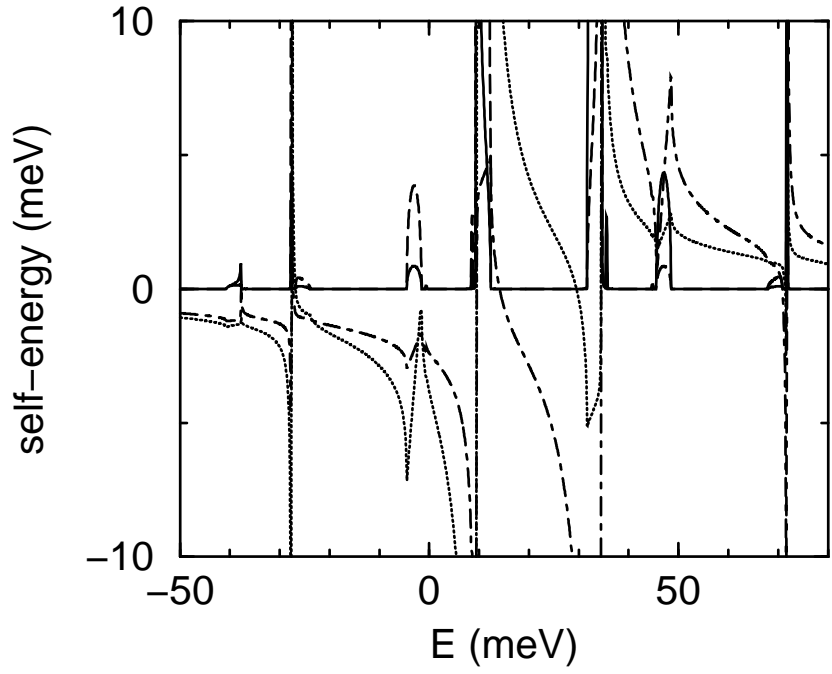}
\caption{Real and imaginary parts of electronic self-energy in the
states with $n=0,1$ for the detuning  $E_{LO}-E_1=-8$\,meV in GaAs at
77\, K. The full and dashed lines denote, respectively, the imaginary
part of the self-energy $(-ImM_0)$ and $(-ImM_1)$, while the dashed
(dotted) line denotes $ReM_0$ ($ReM_1$).}
\label{graf214am}
\end{figure}

The numerical calculations are
performed for GaAs, InAs and InP \cite{Nag,pub37}.
The Dyson equation (\ref{Dyson}) for the retarded self-energy is
solved by iterations.  In contrast to the previous section,  
the Eqs.\ (\ref{Dyson})  
are solved numerically,
substituting a positive finite number $\Delta$ instead of $0_+$ 
in\ (\ref{Dyson}). We used $\Delta=10^{-3}$\,meV. 
The real  and imaginary parts of the self-energy are presented in
Fig.\ \ref{graf214am} 
for GaAs for the detuning chosen to be $E_{LO}-E_{LO}=-8$~meV at
$T_L=77$\,K.
This Figure shows the rather complicated satellite structure of the
electron self-energy obtained in the Tamm-Dancoff approximation. In the
curves, displayed in the Figure, the phonon satellites of both the
electronic ground and excited states are observed.

\begin{figure}
 \epsfxsize=13.5cm
 \epsfbox{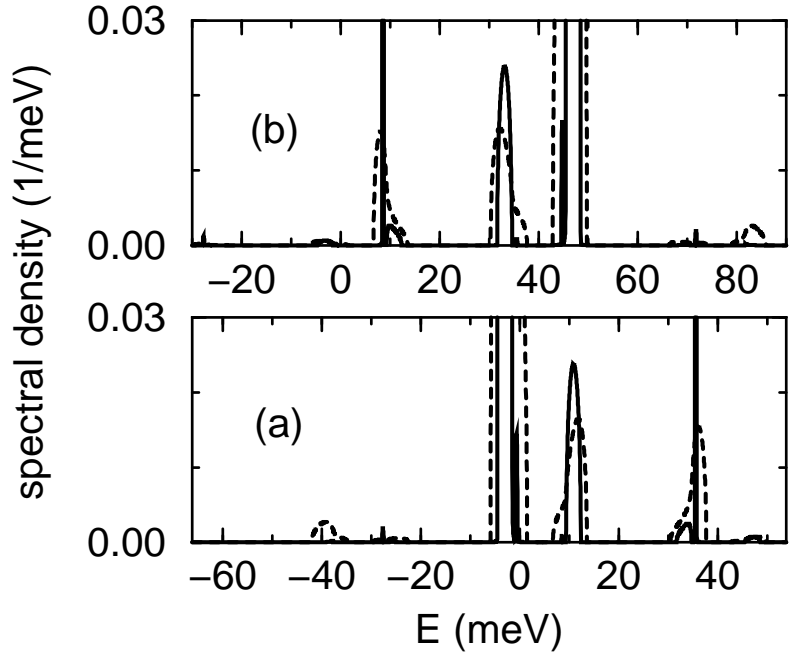}
 \caption{Spectral densities $\sigma_0(E)$ (a) and $\sigma_1(E)$ (b) 
 in GaAs computed for the detuning  $E_{LO}-E_1=-8$\,meV, at the lattice 
 temperature of 
 77\,K 
  (full line) and at the room temperature (dashed line). 
 Note that he $E-$axis of graph (b) is shifted 
 by the energy of optical phonon with respect to
 graph (a).}
 \label{oba}
 \end{figure}

The spectral density of the electron in the excited state would be
$\delta(E-E_1)$ in the noninteracting case ($H_1=0$). 
In the interacting
electron-phonon system the spectral densities $\sigma_0(E)$ and 
$\sigma_1(E)$ have a
rather complex structure displayed in Fig.\ \ref{oba}.
 In this Figure the numerical data
are computed for the parameters of GaAs at the detuning
$E_{LO}-E_1=-8$~meV at $T_L=77$\,K. 
Namely, the
optical-phonon energy  $E_{LO}=36.2$\,meV and the value of  the
 energy 
of the
excited unperturbed electronic state is chosen to be 
44.2\,meV.   Both 
Figs.\ \ref{oba}a
and\ \ref{oba}b show multiple peaks. It is seen that one of the
characteristic energy separations among the individual graphs 
is 36.2\,meV.
This can be interpreted as follows: an  electronic state in the 
quantum dot
with the electron-phonon interaction 
appears to consist of  such components, which  correspond to the two 
unperturbed
electronic states coupled to various numbers of LO-phonons.  
For example, the peak near $E=0$ in Fig.\ \ref{oba}a corresponds 
to the
phonon-less $n=0$ line, while the peak near 36\,meV corresponds to a
one-phonon satellite of the electronic ground state feature 
($n=0$). In the same Fig.\ \ref{oba}a the feature near 8\,meV
can be seen as a one-phonon satellite of the  excited 
state
 ($n=1$), having the energy decreased from $E_1$ by $E_{LO}$. 
 Similar
energy relations are observed in the Fig.\ \ref{oba}b. The main 
peaks in Fig.\ \ref{oba} appear to be spread over an interval of 
several meV. The shape of the peaks was 
characterized in the previous paper
\cite{pub26}, where it was shown, that the shape of the individual
maxima in the pattern of the spectral density on the quantum dot
described by the present model, do not have the form of a Lorentzian. 
Although the present results do not allow to compare with
the known experimental data on luminescence, it may be said that 
the occurence of the
phonon satellites in the spectral densities is in a qualitative 
agreement
with the experimental data on luminescence, in which luminescence 
maxima
separated by one optical-phonon energy use to be  observed (see e. g.
\cite{pub16,pub19,pub20}).

A considerable attention has  been recently   paid in experiment to the 
width of the luminescence lines  \cite{Empe,Banin,Krauss} in
connection with the coherent phonons in the nanocrystals and
semiconductor quantum dots.  The issue of the variation of the "width" 
of a spectral line, whether it decreases or increases, with changing such
parameters like the quantum dot size or the detuning, might help to
judge the relevance of a particular theoretical model of the quantum dot
or a nanocrystal. As it is seen in Fig.\ \ref{oba}, and as it was
discussed using analytical arguments in the reference
\cite{pub26}, the shape of a feature in the spectral density differs
from a Lorentzian, so that it not straightforward to speak about the
linewidth, at least in the case of the present model based on the LO
phonons. An attention should be given to generalizing the present model
in order to include the acoustic phonons and the electron-hole 
interaction. This
question will not be further analyzed in this work.  

A significant property of the spectral densities $\sigma_0(E)$ 
and $\sigma_1(E)$ is
their mutual overlap. Namely, from Fig.\ \ref{oba} it is 
observed that
despite of the 'nearly discrete' structure of these spectral 
densities, 
the
function $\sigma_0(E\pm E_{LO})$ has 
a nonzero overlap with $\sigma_1(E)$, in a rather broad
and continuous range of the detuning $E_{LO}-E_1$. 
This property of the spectral
densities guarranties a nonzero electronic 
relaxation rate in the broad range of the
detuning $E_{LO}-E_1$. 

\begin{figure}
 \epsfxsize=13.5cm
 \epsfbox{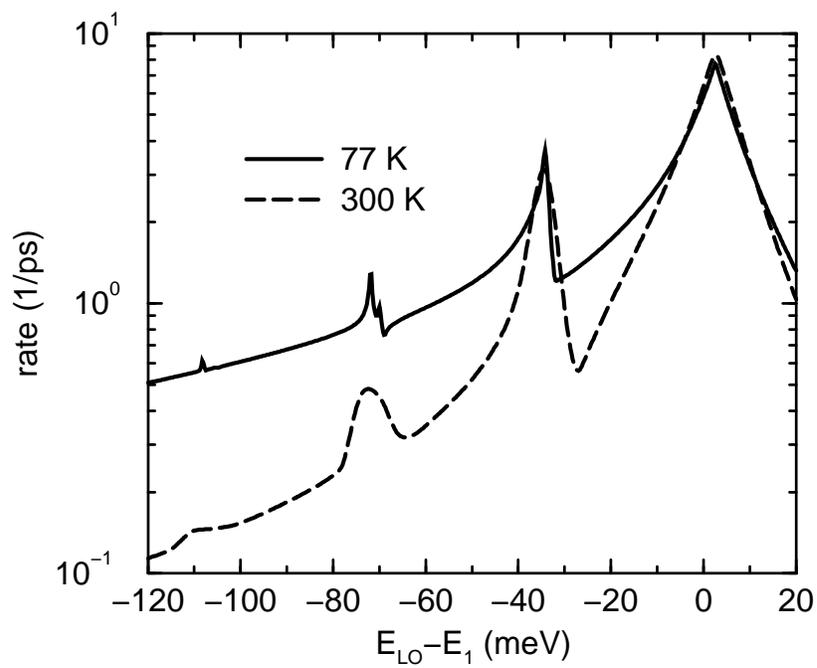}
 \caption{Relaxation rate $(-dN_1/dt)$ in GaAs as a function of the
 detuning $E_{LO}-E_1$ at two temperatures of the lattice.}
 \label{gaas73}
 \end{figure}

The numerical results giving the rate of the
excited state depopulation, 
$dN_1/dt$, as a function of the detuning
$E_{LO}-E_1$, are displayed in Fig.\ \ref{gaas73} for GaAs. 
The relaxation rate is computed for the initial state with 
$N_1=1$ and
$N_0=0$ and with two chosen temperatures of the lattice.
The most intensive relaxation
occurs, when the optical-phonon energy is at resonance with the
excited-state energy $E_1$. In the region of small dots 
($E_{LO}-E_1<0$) the rate
displays an overall decrease with decreasing the dot size,
 at both temperatures. At such a detuning, which
corresponds to $E_1$ equal to $2E_{LO}$, $3E_{LO}$ and 
$4E_{LO}$ ($E_{LO}-E_1$ equal to about
$-$36\,meV, $-$72\,meV and $-$108\,meV, respectively) further resonance 
maxima appear. 
Although the sharp peaks of the spectral densities, as they 
are displayed in
Fig.\ \ref{oba}, broaden when the lattice temperature is increased, 
the overall
values of the relaxation rate decrease with increasing 
 temperature. This decrease of the electronic relaxation rate  with
 increasing the temperature of the lattice can be explained taking into
 account the nature of the electronic states determined  
 in the self-consistent
 Tamm-Dancoff approximation to the electronic self-energy. The electron
 is coherently coupled to a number of the LO-phonons. This property of
 this electron-phonon system is responsible for the effect of a non-zero
 relaxation even at the condition of nonzero detuning $E_{LO}-E_1$. As
 the temperature of the lattice increases, this coherence weakens, which
 leads to a decrease of the relaxation rate. 
 
\begin{figure}
 \epsfxsize=13.5cm
 \epsfbox{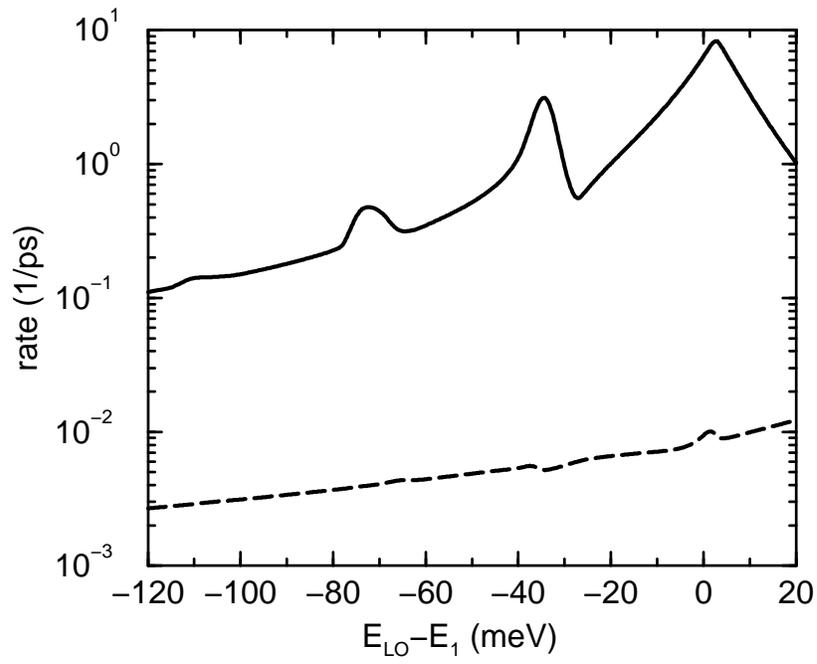}
 \caption{Two contributions to the total relaxation rate $(-dN_1/dt)$ 
  in GaAs at $T_L=300$\,K. The full line gives the contribution
 of the process accompanied by the emission of LO phonon, while the
 dashed line gives the one of the process with LO phonon absorption.}
 \label{gaas300m}
 \end{figure}

It may be interesting to see, how much the process accompanied by the
absorption of the LO-phonon (the term proportional to
$N_1(1-N_0)\nu_{LO}$ in eq. (18)) contributes to the total relaxation
rate at $t=0$.  The contribution of the latter term 
is given by the dashed line in Fig.\ \ref{gaas300m}, while the
contribution of the main term, the one proportional to $N_1(1-N_0)(1+
\nu_{LO})$, is given by the full line. 
We see, that in accord
with the fact that the electron-LO-phonon coupling in GaAs is rather
weak, the phonon-absorption term 
gives only a rather small contribution to the total rate
$dN_1/dt$.

The relaxation rate  
reaches the maximum of the order of about 1 electron per 
100 femtoseconds.
This value  has the order of magnitude of the experimental
observations in some studied samples \cite{pub17}, 
in which the electronic relaxation
rate was characterized by the relaxation time of 
 several hundreds of fs. 
The relaxation rate computed at the room
temperature of the lattice, at which the relaxation rate is decreased
with respect to the low temperature case,  
is still at the scale of tens of picoseconds. 
These theoretical
results, obtained within the present simple model of the quantum dot,
  therefore 
 contradict   the assumption of  an existence of the phonon
bottleneck in the electron-energy relaxation in quantum dots 
in polar
semiconductors, at least in the case of small dots, such, in which the
detuning is near to zero or negative. 
Let us remind, that the present theory is formulated 
for small dots (negative detuning). 
For large dots ($E_{LO}-E_1>0$) the present model 
has to be  extended
to include the effect of more unperturbed electron energy levels. This
generalization of the model of the quantum dot is not done in this work.

\begin{figure}
 \epsfxsize=13.5cm
 \epsfbox{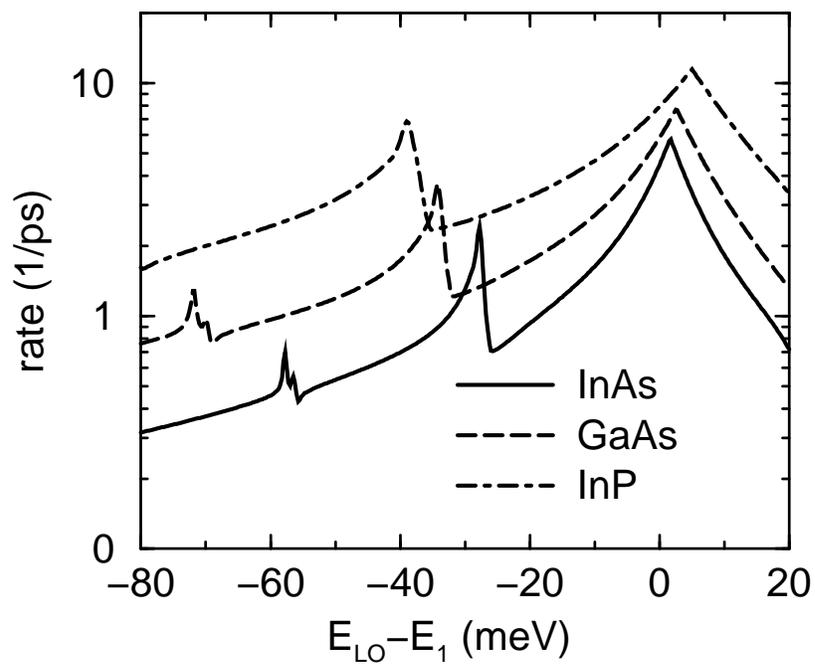}
 \caption{A comparison of the relaxation rate $(-dN_1/dt)$ at
 $T_L=77$\,K in three semiconducting materials.}
 \label{spolu}
 \end{figure}

The overall efficiency of the presently considered mechanism of the
electron energy relaxation depends on the strength of the Fr\"ohlich
coupling. This dependence is presented in Fig.\ \ref{spolu}, 
in which the
relaxation rate is displayed for InAs, GaAs and InP at 77\,K of the
lattice temperature.  

Summing up, using a single-electron model of quantum dot with two
electron energy levels and with the electron coupled to dispersionless
bulk optical phonons, this work
 presents 
the calculation of 
the electron-energy relaxation in quantum dots in polar
semiconductors. The relaxation time is found to be 
in the  interval from hundreds of
femtoseconds
to tens of picoseconds in such quantum dots, in a broad range of the
detuning, in which the 
energy of the 
electronic excited
state is equals from about one LO-phonon energy 
to about 
three or four times the 
energy of the LO-phonon. This overall agreement
of the present results with the theoretical data suggests, 
that the electron-LO-phonon
interaction  provides  an effective multiphonon mechanism 
giving the 
femtosecond or picosecond
electron-energy relaxation time
in a rather broad range of the quantum dot size of small quantum dots.
 Such
conclusions apply to the temperature range from low temperatures of the
lattice to the room temperature. 

\begin{appendix}
\section{Equilibrium correlation functions}
In this Appendix several formulas are reminded, connecting the
correlation functions with the spectral density and the self-energy in
an equilibrium system at nonzero temperature. 
In particular, an
approximate formula expressing the time dependent correlation functions
in terms of the self-energy, suitable for a generalization to the case
of a nonequilibrium system, is presented. 

In the thermodynamic equilibrium at temperature $T$
the time-dependent single-particle correlation functions can be
spectrally decomposed in the well-known way \cite{pub31} according to
the formulas:
\be
<B(t')A(t)> =\frac{1}{2\pi}\int^{\infty}_{-\infty} J_{BA}(\omega)
e^{i\omega(t'-t)}d\omega
\ee
and 
\be
<A(t)B(t')>=\frac{1}{2\pi}\int^{\infty}_{-\infty} J_{BA}(\omega)
e^{\frac{\hbar\omega}{k_BT}} e^{i\omega(t'-t)}d\omega, \label{ctverec}
\ee
where
$J_{BA}(\omega)$ is spectral intensity of the correlation function 
$<B(t')A(t)>$. The spectral intensity can be obtained from the knowledge
of the retarded thermodynamic Green's function 
\be
\ll A(t);B(t') \gg _r=\frac{1}{i\hbar}\theta(t-t')
<[A(t),B(t')]_{\eta}>,
\ee
where the symbol $<\dots>$ denotes the averaging  performed over 
the grand-canonical ensemble,
$A(t)$$=$$\exp(i{\cal H}t/\hbar) A \exp(-i{\cal H}t/\hbar)$, ${\cal H} =
H-\mu N$, 
 $\mu$ is the chemical potential and $N$ is operator of number of
particles. Also, $[A,B]_{\eta}=AB-\eta BA$, in which $\eta=\pm 1$ for
Bose and Fermi particles, respectively. Defining the Fourier
picture $\ll A;B\mid \omega \gg_r$ as
\be
\ll A(t);B(t')\gg_r=\frac{ 1}{2\pi}\int^{\infty}_{-\infty}
e^{-i\omega(t-t')}d\omega \ll A;B\mid\omega \gg_r
\ee
and when the spectral intensity is real, then there is the relation
between the Green's function and the spectral intensity
\be
Im\ll A;B\mid\omega\gg_r=-\frac{1}{2\hbar}\left(
e^{\frac{\hbar\omega}{k_BT}}-\eta\right)J_{BA}(\omega). \label{si}
\ee

Assuming in the present work that the influence of the electron-phonon
interaction on the characteristics of the phonons is only minor, the
phonon correlation functions will be determined in the zero-order
approximation. Realizing that \cite{pub31}
\be
\ll b_{\bf q};b^+_{\bf q}\mid\omega\gg_r= \frac{1}{
\hbar\omega-E_{LO}+i0_+},
\ee
one obtains
\be
<b_{\bf q}(t_1)b^+_{\bf q}>+ <b^+_{\bf q}(t_1)b_{\bf q}>
=(\nu_{LO}+1)e^{-i\omega_{LO}t_1}+\nu_{LO}e^{i\omega_{LO}t_1},
\label{corrfon}
\ee
with 
\be
\nu_{LO}=\frac{1}{e^{\frac{E_{LO}}{k_BT}}-1}.
\ee

The electronic time-dependent correlation functions will be calculated
with nonzero electron-phonon coupling $H_1$. In this case the spectral
intensity will be expressed in terms of the electronic self-energy. In
the diagonal approximation neglecting the inter-level correlations due
to the electron-phonon coupling, the retarded Green's function of the
$n$-th electronic state can be written as
\be
\ll  c_n;c^+_n\mid \omega \gg _r = \frac{1}{ \hbar\omega -E_n+\mu
-M_n^{(r)}(\hbar\omega)+i0_+},
\ee
where $M^{(r)}_n(\hbar\omega)$ is the retarded self-energy of the $n$-th
electronic state. Working with the retarded self-energy only we shall
drop the index $(r)$ in the symbol $M^{(r)}_n(\hbar\omega)$. 
Defining $R_n(\hbar\omega)=ReM_n(\hbar\omega)$
we shall write $M_n(\hbar\omega)=R_n(\hbar\omega)-iY_n(\hbar\omega)$.
Such a decomposition of the self-energy into the real and imaginary
parts allows one to work with the real and non-negative
 quantity $Y_n(\hbar\omega)$ in what follows.
 

With (\ref{ctverec}) the correlation function $<c_n(t_1)c^+_n>$ reads:
\be
<c_n(t_1)c^+_n>= \frac{\hbar}{\pi}\int^{\infty}_{-\infty}
\left(1-\frac{1}{e^{\frac{\hbar\omega}{k_BT}} +1}\right)
\frac{Y_n(\hbar\omega)+0_+}{(\hbar\omega- E_n+\mu 
-R_n(\hbar\omega))^2+
(Y_n(\hbar\omega)+0_+)^2} e^{-i\omega t_1}d\omega. \label{2cross}
\ee
Let us notice, that in the zero-order approximation, in which $R_n$ and
$Y_n$ go to zero, the expression $\exp (\hbar\omega /k_BT)$ in the
latter formula can be substituted by $\exp ((E_n+\mu)/k_BT)$.
In formula (\ref{2cross}) we shall approximately substitute the
expression $1/(\exp (\hbar\omega/k_BT)+1)$ by the value of the
Fermi-Dirac distribution function, giving the average number of
electrons in the $n$-th electronic state. In this work therefore the
electronic population will be approximated by
\be
N_n=\frac{1}{e^{\frac{E_n-\mu}{k_BT}}+1}.
\ee
Two correlation functions then are:
\be
<c_n(t_1)c^+_n>=\hbar (1-N_n)\int^{\infty}_{-\infty}\sigma_n(\omega)
e^{-i\omega t_1}d\omega, \label{corrf1}
\ee
\be
<c_n^+(t_1)c_n>=\hbar N_n\int^{\infty}_{-\infty}\sigma_n(\omega)
e^{i\omega t_1} d\omega,  \label{corrf2}
\ee
where the spectral density $\sigma_n(E)$ is
\be
\sigma_n(\omega)=\frac{1}{\pi}\frac{Y_n(\hbar\omega)+0_+}{ 
(\hbar\omega -E_n+\mu
-R_n(\hbar\omega))^2+ (Y_n(\hbar\omega)+0_+)^2},  \label{shu}
\ee
going to the $\delta$-function in the limit of zero self-energy and
  fulfiling the rule
\be
1=\hbar \int^{\infty}_{-{\infty}}\sigma_n(\omega)d\omega,
\ee
for $n=0,1$. 

The formulas (\ref{corrfon}), (\ref{corrf1}) and 
(\ref{corrf2}) provide the dependence of the spectral densities on the
temperatures, or on the populations, 
of the electrons and phonons, within the theory of the real time
thermodynamic Green's functions.
These functional dependences will be assumed to be valid 
for the correlation
functions, which we meet in the formula (\ref{scheme}) 
in the general case
of nonequilibrium state of the system.
In this way, 
the electronic relaxation rate (\ref{rate}) 
 can  be expressed in terms of the electronic spectral density and in
 terms 
of the electronic self-energy.

The above given method of obtaining the single-particle correlation
functions in the nonequilibrium system under study is accompanied by an
inaccuracy, which may depend on  the strength of the electron-phonon
coupling. The reason for this is obvious upon comparing the statistical
operators,
with the help of which the averaging is performed. In the
case of the thermodynamical equilibrium the averaging is performed in
the grandcannonical ensemble of the system with the electron-phonon
interaction included, while in the averages like
$<c^+_i(t)c_j>_q$, appearing in the formula (\ref{scheme}), the
quasi-equilibrium statistical operator does not depend on the
electron-phonon interaction, at least 
in the approximation we assumed in this
work. This question can be clarrified with the use of a more systematic
approach to this nonequilibrium system.
\section{Electronic self-energy}
Using the language of the Feynman diagrams for the Matsubara Green's
functions \cite{pub34,pub35},
 the self-consistent Tamm-Dancoff
approximation (or the self-consistent Born approximation) 
to the electronic self-energy can be expressed as a
diagram containing two bare interaction vertexes connected with two lines,
one of them corresponds to the full electronic Green's function $G$, while
the other corresponds to the bare phonon Green's function $D^{(0)}$. 
The equation for the
self-energy of the electron in the $n$-th state, $M_n(i\hbar \omega_p)$,
defined on the set of imaginary frequencies $\omega_p=(2n+1)\frac{\pi
k_BT}{\hbar}$, $n$ is integer, is
\bea
 M_n(i\hbar\omega_p)=  k_BT\sum_m \sum_{\bf q} \mid A_q\mid
^2\mid \Phi(m,n,{\bf q})\mid ^2 \label{td} 
 \sum_r G_m(i\omega_p-i\omega_r)D^{(0)}(i\omega_r). 
\eea 
Using the Lehmann's representation of the Green's functions \cite{Mahan}, 
the summation 
 over the discrete imaginary frequencies 
 $\omega_r=2r(\pi k_BT/\hbar)$, $r$ being an integer, can be
performed.
 In this way the 
  equation for the retarded electronic self-energy in the
 state $m$ of
 is obtained:
\bea
 M_n(E)= 
 \label{i} \sum_m
\alpha_{nm}[G^{ret}_m(E-E_{LO})(1-n_F(E-E_{LO})+\nu_{LO})\\
+G^{ret}_m(E+E_{LO})(n_F(E+E_{LO})+\nu_{LO})], \nonumber 
\eea
where $n_{F}(E)=1/(\exp(E/K_BT_e)+ 1)$ and
$\nu_{LO}=1/(\exp(E_{LO}/k_BT_{L})-1)$  
Here  $G^{ret}$ denotes the retarded electronic Green's function. 
Note, that in the latter equation the electronic Fermi-Dirac
distribution function  $n_{F}(E)$  appears
as a function of the energy variable $E$. Assuming the knowledge of the
electronic temperature, this function would present no difficulty.
However, 
in
this work we approximate these electronic distributions 
 by the Fermi-Dirac distribution function value $N_m$, which is the
 value of the Fermi-Dirac distribution function  
in the state with the unperturbed electronic energy $E_m$, at which the
corresponding retarded Green's function in the latter equation (\ref{i}) would
have the pole in the case of no electron-phonon coupling. In this way
the treatment of the electronic temperature can be avoided.

The equation for the retarded
self-energy is  ($E_{LO}=\hbar\omega_{LO}$):
\bea
\lefteqn{ M_n(E)=\sum_m\alpha_{nm} \left[ \right. 
 \frac{1-N_m+\nu_{LO}}
 {E-E_m-E_{LO}-M_m(E-E_{LO})+i0_+}} \nonumber \\
 +&&\frac{N_m+\nu_{LO}}{E-E_m+E_{LO}-M_m(E+E_{LO})+i0_+} 
\left. \right] . \label{finaltdB}
\eea
The  equation (\ref{finaltdB}) closely resembles 
the
similar equation for the self-energy of Frenkel excitons studied earlier
\cite{pub36}. In the reference \cite{pub36} this equation is derived
with help of the real-time thermodynamic Green's functions.

In (\ref{finaltdB}) the chemical potential is not
written. This form of the equation corresponds to the choice of the
Hamiltonian $H$ in the equation of motion 
for the Green's function, instead of the
Hamiltonian ${\cal H}=H-\mu N$, although the statistical averaging is
still performed with the grand-canonical ensemble. Consistently with
this choice of Hamiltonian, the chemical potential 
$\mu$  is  dropped in eq. (\ref{shu}).
\end{appendix}

%
%
%
%
%
%
%
%
%

\end{document}